\documentclass[letter]{aa} 

\usepackage{graphicx}
\usepackage{txfonts}
\begin{document}

   \title{Mass-ratio distribution of extremely low-mass\\ white dwarf binaries}

\titlerunning{Mass ratio distribution of ELM}

 \author{Henri M.J. Boffin}

   \institute{ESO, Alonso de C\'ordova 3107, Casilla 19001, Santiago, Chile; 
              \email{hboffin@eso.org} }

\date{Received January 28, 2015; accepted February 15, 2015}

\abstract{Knowing the masses of the components of binary systems is very useful for constraining the possible scenarios that could lead to their existence. While it is sometimes possible to determine the mass of the primary star,  it is challenging
to obtain good mass estimates of the secondary of a single-line
spectroscopic binary. If the sample of such binaries is large enough, however, it is possible to use statistical methods to determine the mass-ratio distribution, and thus, the mass distribution of the secondary. Recently, the mass distribution of companions to extremely low-mass white dwarfs was studied using a sample of binaries from the ELM WD Survey. I reanalyse the same sample with two different methods: in the first one, I assume some functional form for the mass distribution, while in the second, I apply an inversion method. I show that the resulting companion-mass distribution can be as well approximated by either a uniform or a Gaussian distribution. The mass-ratio distribution derived from the inversion method without assuming any a priori functional form shows some additional fine-grain structure, although, given the small sample, it is difficult to claim that this structure is statistically significant. I conclude that it is not yet possible to fully constrain the distribution of the mass of the companions to extremely low-mass white dwarfs, although it appears that the probability to have a neutron star in one of the systems is indeed very low.} 


\maketitle
%

\section{Introduction}
Extremely low-mass white dwarfs (ELMs) are thought to be the end products of binary star interactions and generally have a companion \citep[e.g.,][]{Marsh1995,2011MNRAS.413.1121R}. Although still far from understood,  binary models predict that the companions to these ELMs are white dwarfs because the system has undergone one or two common-envelope phases \citep{2001A&A...365..491N,2006A&A...460..209V,Woods2012,2014A&A...562A..14T}. Knowing the distribution of the companion's mass to these ELMs could provide useful constraints on the various parameters that
enter binary star evolution and the common-envelope phase, as well as to predict whether and when they will merge and, more generally, how they will evolve \citep{2012MNRAS.422.2417D,2012ApJ...751..141K}. Thanks to the ELM Survey, which has identified 61 ELMs and provided orbital parameters for 54 of them\footnote{\citet{ori} used the 55 systems  of \citet{Gianninas2014}, which are essentially based on the list of 54 systems from \citet{Brown2013}. Here I used these 54 latter systems because they come from a homogeneous sample.} \citep[see][and references therein]{Brown2013}, it is now possible to begin envisaging conducting a statistical analysis as was done by \citet{ori}. The problem is that ELMs are single-lined spectroscopic binaries and as such, it is not possible to obtain the mass ratio directly \citep[see, e.g.,][]{2010A&A...524A..14B,2012ocpd.conf...41B,Cure2014}, but one needs to apply statistical methods to derive the mass-ratio (or companion mass) distribution. \citet{ori} developed a Bayesian probabilistic model to infer the companion mass distribution for the above-mentioned sample, {\it \textup{assuming a functional form}} -- a two-component Gaussian, with one component representing white dwarfs with masses between 0.2 and 1.44~M$_\odot$, and the other neutron stars with masses centred around 1.4~M$_\odot$ and a standard deviation of 0.05~M$_\odot$. Using a Markov chain Monte Carlo algorithm, they found that their best fit is given by a population of white dwarfs centred around 0.74~M$_\odot$ and a standard deviation of 0.24~M$_\odot$, without a neutron star. This is quite an interesting result that also indicates that in contrast to population synthesis models, the majority of companions to ELMs are CO-core WDs, and not another He WD. As such, it is important to examine whether this result holds when using different methods, including when the functional form is not fixed a priori. This is what I present here.

\section{Methods}

The derivation of the orbital elements for a single-lined spectroscopic binary (period, radial velocity amplitude, and eccentricity) allows obtaining the spectroscopic mass function, $f(m)$, which is a combination of the masses of the two components and the (unknown) inclination of the orbit on the line of sight, $i$: 
$$ f(m) = \frac{M_2^3}{(M_1+M_2)^2}~\sin^3  i ,$$
where $M_1$ is the mass of the primary (in this case, the ELM), and $M_2$, the mass of its companion. If $M_1$ is known, as it is the case here, then one can rewrite this as a function of the mass ratio, $q=M_2/M_1$:
$$ Y= \frac{f(m)}{M_1} = \frac{q^3}{(1+q)^2}.$$
The distribution of the logarithm of $f(m)$ -- or $Y$ --  can
be used to determine the distribution of $M_2$ or  $q$. This was done here using two different methods, in which I always assumed that the inclination $i$ is randomly distributed on the sky, that is, $P(i)=\sin i$. I used the sample of \citet[][similar to that used by \citet{ori}, but see the footnote]{Brown2013}, which provides a list of 54 systems with known $f(m)$ and $M_1$. 

In the first method, I assumed a functional form for the distribution of $M_2$: this is either a Gaussian with mean $\mu$ and standard deviation $\sigma$, or a uniform distribution, defined between a lower, $M_{\rm 2,l}$, and an upper, $M_{\rm 2,u}$ value of the companion mass. I then applied a Kolmogorov-Smirnov statistical test. 
I assumed a single population of companions, without distinguishing between a neutron star (NS) and a white dwarf companion population,
for instance. This is based on the fact that \citet{ori} found the NS fraction to be very low, as I confirm here as well.

In each case, I ran 10,000 Monte Carlo simulations, where $M_1$ is distributed according to the observed distribution (see the
online Fig.~\ref{fig:m1}), $M_2$ according to the chosen distribution with one set of parameters, and $i$ is assumed to be randomly distributed on the sky. For each sample of simulations, I calculated the cumulative distribution of $\log f(m)$, which I compared with the observed one. To do this, I calculated the largest deviation between the two distributions, $D^*=D_{54,10000}$. Running 10,000 simulations provides very good precision, and there would be no gain in running more, as the estimator $D^*$ saturates for large numbers\footnote{This is because $D_{\rm n,n'} \propto \sqrt{\frac{n+n'}{ n  n'}} \propto \sqrt{\frac{1}{n}}$ for  $n'>>n$ and $n>>1$.}.

This estimator allows determining the probability with which  the simulated and the observed distribution are extracted from the same population. Thus, if $D^*>0.2628$,  the chance is only 0.1\% that the two populations are extracted from the same population, and we may most likely ignore such solutions. For $D^*>0.1274$ and $D^*>0.0506$, these probabilities become 33\% and 99.9\%, respectively. The first value provides a 1-$\sigma$ estimate of the parameters that are allowed, while the latter number can be used to estimate the 0.1\% confidence interval of parameters that provide a very good match to the observed distribution.

In the second method, I took a more direct approach and inverted the distribution of $Y$ to derive the mass-ratio distribution.

\section{Results}

\subsection{Functional form}

\subsubsection{Gaussian}

\begin{figure}[htbp]
   \centering
   \includegraphics[width=6cm]{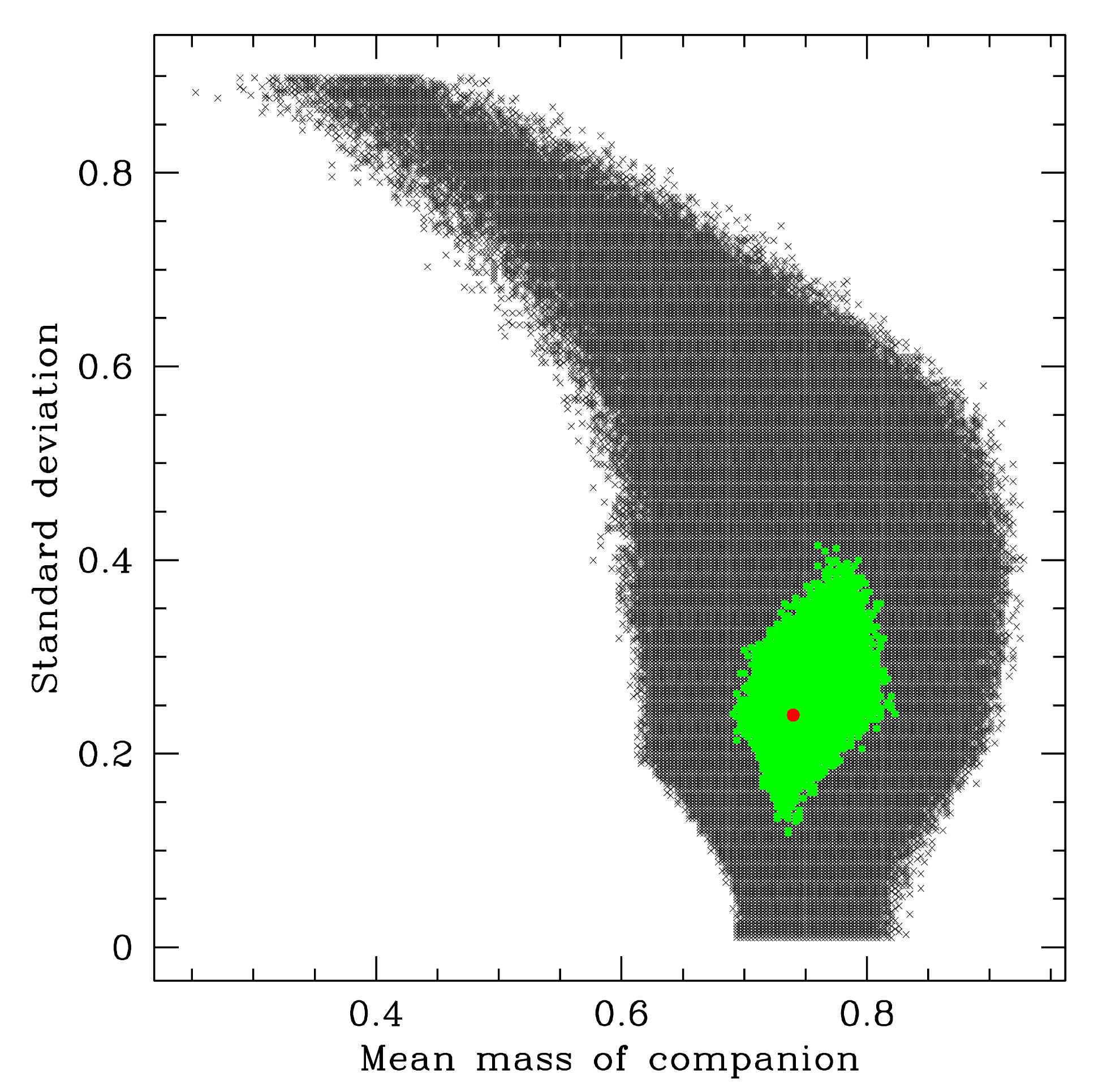} 
   \caption{Standard deviation, $\sigma$, versus the mean mass of the companion, $\mu$, for all simulated samples where the companion mass is distributed according to a Gaussian, and which have $D^*< 0.1274$ (in black) and those that have $D^*< 0.0506$ (in green).  The heavy, red dot shows the location $\mu=0.74$, $\sigma=0.24$.}
   \label{fig:gaussM2}
\end{figure}
 
As mentioned above, I here assumed that the distribution of the companion mass, $\Phi(M_2)$, is given by a Gaussian:
$$ \Phi(M_2) \propto \exp(-\frac{M_2-\mu}{2 \sigma^2}),$$ and I determined for a plane $0 < \mu< 2, 0 < \sigma < 1$ the value of $D^*$. The relation of $D^*$ with $\mu$ is shown in the online Fig.~\ref{fig:gaussD}, where I also show the lines corresponding to $D^*=0.0506,$ 0.1274, and 0.2628. This figure shows that one can find simulations spanning the whole range $0 < \mu < 1.22~{\rm M}_\odot$ that would lead to 
$D^*<0.2628$, while the 1-$\sigma$ range covers the values  $0.3 < \mu < 0.92~{\rm M}_\odot$ . This indicates that it is difficult to constrain the parameters of such a functional form based on the small observed sample. 
It is clear, however, that the simulations with $\mu$ between 0.7 and 0.8~M$_\odot$ correspond to the lowest values of $D^*$, with the minimum being around $\mu$=0.76~M$_\odot$ and $\sigma$=0.27~M$_\odot$, leading to $D^*=0.027$. This value indicates that this simulation and the observed distribution are consistent at a very high level
because the null hypothesis that the two populations are drawn from the same population can only be rejected at a level of $1.3 \times 10^{-12}$\%. 

Of course, the standard deviation $\sigma$ is correlated with $\mu$, and Fig.~\ref{fig:gaussM2} shows the solutions that lead to $D^*<0.0506$ and $D^*<0.1274$, illustrating the allowed range. For $D^*<0.1274$, I found that the mean value of the companion mass is $\mu_m = 0.70 \pm 0.12~{\rm M}_\odot$, while the mean value of the standard deviation is $\sigma_m=0.45\pm 0.22~{\rm M}_\odot$. For $D^*<0.0506$, these values become $\mu_m = 0.76 \pm 0.03~{\rm M}_\odot$ and $\sigma_m=0.27\pm 0.06~{\rm M}_\odot$. 

A $\chi^2$ analysis that computes the deviation of the computed and observed distribution of $\log f(m)$ provides a similar result, with the best fit being given by $\mu = 0.76 \pm 0.02 ~{\rm M}_\odot$ and $\sigma = 0.27\pm 0.02~{\rm M}_\odot$ (see the online Fig.~\ref{fig:x2map}).

\begin{figure}[tbp]
   \centering
   \includegraphics[width=6cm]{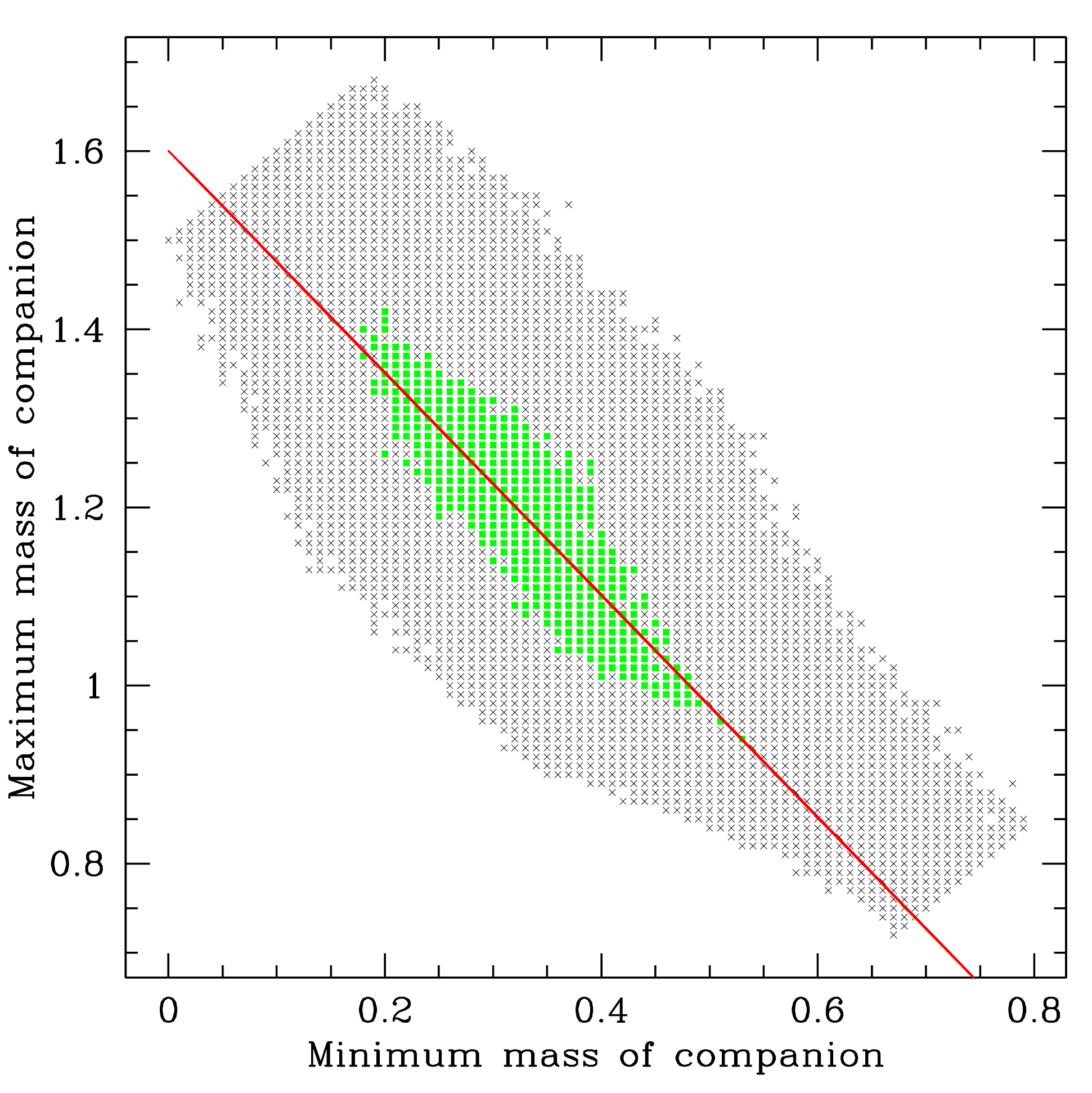} 
   \caption{Highest mass of the companion as a function of the lowest mass of the companions for all simulated uniformly distributed samples that have $D^*< 0.1274$ (in black) and those that have $D^*< 0.0506$ (in green). The two masses are clearly correlated, and I show a linear fit to the green dots as the heavy, solid red line.}
   \label{fig:uniMminMmax}
\end{figure}

\subsubsection{Uniform distribution}
I repeated the same analysis but using a uniform distribution of the companion mass, between a lowest ($M_{2,\rm l}$) and a highest value ($M_{2,\rm u}$), which were assumed to be in the range 
$0 < M_{2,\rm l} < 0.9$~M$_\odot$ and $M_{2,\rm l} < M_{2,\rm u} < M_{2,\rm l}+1.5$~M$_\odot$. The results are shown in the
online Fig.~\ref{fig:uniD} and in Fig.~\ref{fig:uniMminMmax}. Again, the range of allowed values is  very wide. If this is
restricted  to $D^* < 0.2628$, the whole range of  $M_{2,\rm l}$ is allowed, while for $M_{2,\rm u}$, it is restricted to values between 0.5 and 1.95~M$_\odot$. The range becomes narrower for lower values of $D^*$, as shown in Fig. 2, which also shows
that the acceptable values of $M_{2,\rm l}$ and $M_{2,\rm u}$ are correlated.  A linear fit gives 
$$ M_{2,\rm u} = -1.25 M_{2,\rm l} + 1.60. $$

The mean values of $M_{2,\rm l}$ and $M_{2,\rm u}$ are $0.36\pm0.18$~M$_\odot$ and $1.20\pm0.23$~M$_\odot$ for $D^*<0.1274$, and $0.33\pm0.07$~M$_\odot$ and $1.19\pm0.10$~M$_\odot$ for $D^*<0.0506$. The lowest value of $D^*$ (0.02793) is reached for $M_{2,\rm l}=0.25$~M$_\odot$ and $M_{2,\rm u}=1.28$~M$_\odot$. This value corresponds to a probability that the two distributions are  not extracted from the same population of $4 \times 10^{-10}$\%. Perhaps most importantly, this shows that a uniform distribution of the companion mass fits the data very well and cannot be discarded. 
With such a functional form, a Gaussian or a uniform distribution
cannot be preferred; they do, of course, quite overlap (see the
online Fig.~\ref{fig:distmass}).

\begin{figure}[tbp]
   \centering
   \includegraphics[width=9cm]{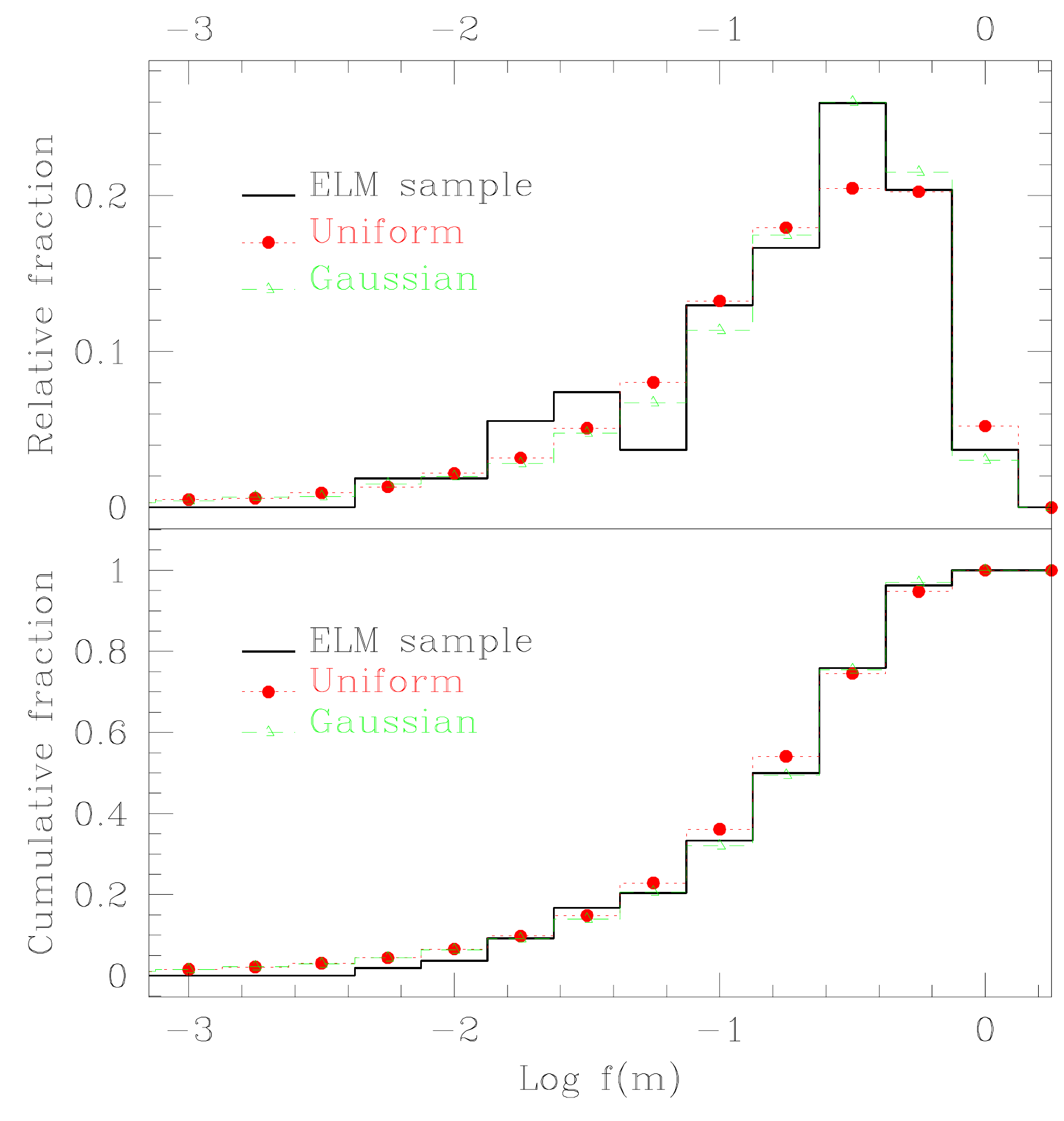} 
   \caption{Comparison between the observed distribution of the logarithm of the spectroscopic mass function (solid black line) and the best fits for the uniformly distributed (red dotted line connected by heavy dots) and Gaussian-distributed (green dashed line connected by open squares) companion masses. For the former, I used $M_{2,\rm l}=0.25$~M$_\odot$ and $M_{2,\rm u}=1.28$~M$_\odot$, while for the latter, I used $\mu = 0.76$~M$_\odot$ and $\sigma=0.27$~M$_\odot$. The top panel shows the fraction of systems, while the bottom panel is the cumulative fraction of systems. It is clear that both samples are good fits, given the intrinsic errors of the observed distribution. }
   \label{fig:logf}
\end{figure}

This is further illustrated in Fig.~\ref{fig:logf}, which shows the distribution of $\log f(m)$ for the observed sample as well as for the two best fits of the functional form, a Gaussian, and a uniform distribution. It is clear that both simulated distributions are good fits to the observed distribution, while it is very hard to distinguish the results from the two different functional forms. 

\subsubsection{Inversion method}

Using a functional form for the mass distribution allows a better control of the systematics of a method, but at the cost of risking missing some interesting deviations. This is for example the case in the paper by \citet{ori} in their test 4, where they compare the distribution they obtain for a sample of post-common envelope systems to that obtained from spectroscopy. Although they clearly reproduce the bulk distribution, they miss the tail and other details of the distribution (see their Fig. 3). Moreover, \citet{ori} also pointed out that for the sample of ELMs, their result ``could indicate that the true WD distribution may not be exactly Gaussian'' -- and indeed the previous section has shown that a uniform distribution also provides a good fit to the observations. It is therefore useful to consider exploring methods that do not require the a priori input of a functional form. This is the case of the Richardson-Lucy (R-L) inversion method, as used by \citet{1993A&A...271..125B,2010A&A...524A..14B,2012ocpd.conf...41B}, and \citet{Cure2014}.
 
\begin{figure}[htbp]
   \centering
   \includegraphics[width=9cm]{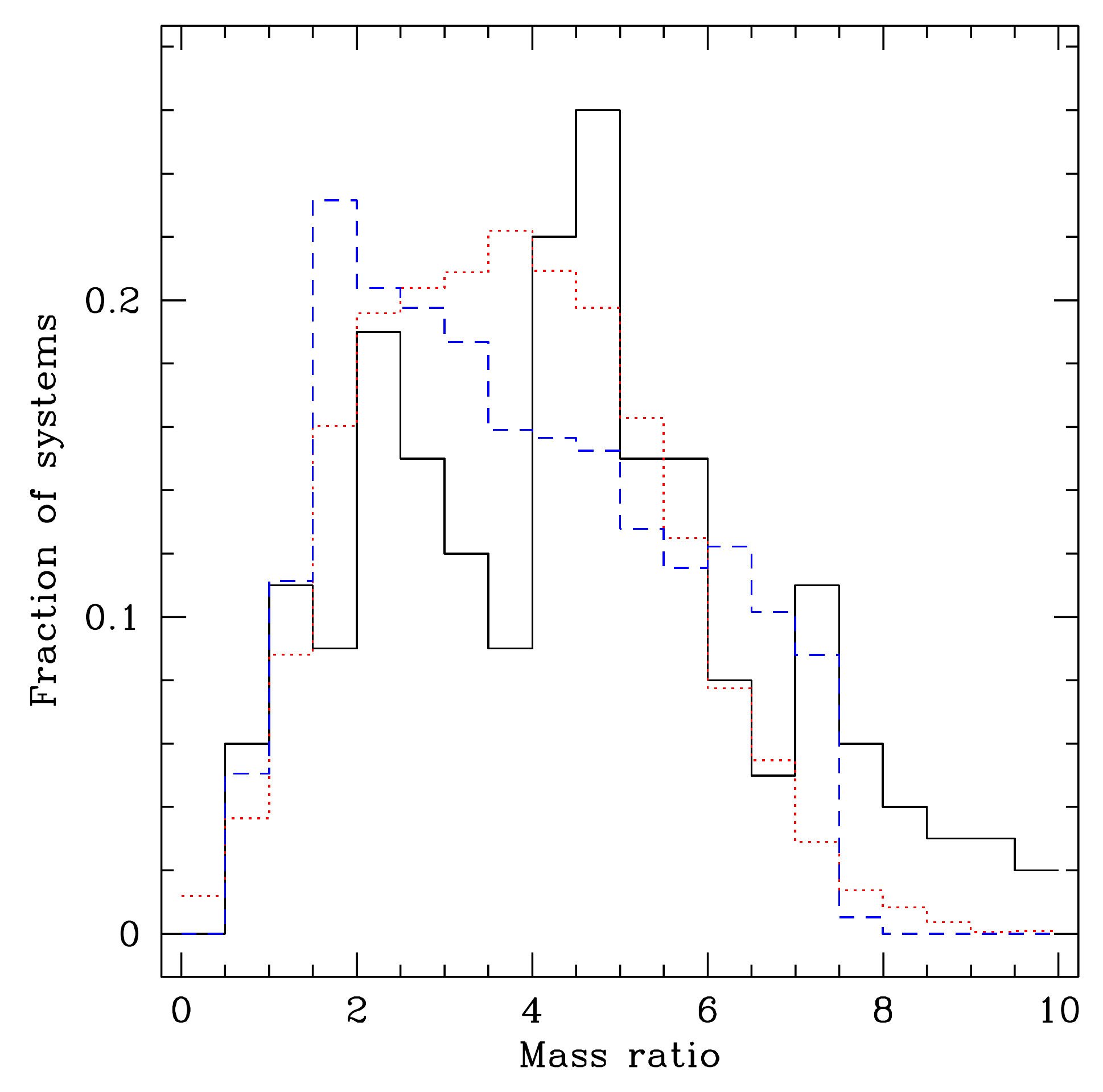} 
   \caption{Distribution of the mass ratios as determined by the Richardson-Lucy algorithm (solid black line) and as obtained for the best fits with a Gaussian  (red dotted line) and a uniform (blue dashed line) distribution of companions.}
   \label{fig:massratio}
\end{figure}

I refer to these papers and references therein for a full discussion of the method,  and in particular to \citet{CB1994} for a more formal presentation. Here, I just mention that the Richardson-Lucy method relies on the Bayes theorem on conditional probabilities and solves the Fredholm integral equation that links $Y$ with $q$ by an iterative scheme. It is 
important to note that we do not directly have the distribution of the companion mass, but only of the mass ratio. This is, however, a very important parameter for binary evolution models, and given that the mass of the primary is very peaked at 0.17~M$_\odot$, the distribution of the mass ratio can give an idea of the distribution of the companion mass. The outcome of this method is shown in Fig.~\ref{fig:massratio}, where I also compare it with the distribution derived from the best fits of the functional form method.  For the latter, I derived the mass-ratio distribution by using the functional form for the companion mass and the primary mass distribution as determined by \citet{Brown2013}.

All three methods provide a rather similar mass-ratio distribution, with some small differences. The outcome of the R-L method gives a broad Gaussian distribution, but with two more pronounced peaks, around $q\sim2-2.5$ and around $q\sim 4-5$. 
This  may indicate that a simple functional form may be missing on some small structure in the data, although a much larger sample
is needed to be able to confirm these \citep{2012ocpd.conf...41B}.  A K-S test indicates that the hypothesis can be rejected that the functional forms and the outcome of the R-L method are drawn from the same population at the level of 92\% -- a rather high number, but perhaps not convincing enough. Indeed, given the small sample, this is at most a two-sigma result. However, it shows that the true mass ratio distribution (and companion mass distribution) {\it may} have a more complicated structure than any simple functional form we can think of. Only with much larger samples will be able to know this.

At the suggestion of the referee, I have also examined the mass-ratio distribution derived with the R-L method when limiting to the least massive primary stars, that is, the 33 systems with masses lower than 0.2~M$_\odot$. The resulting distribution is shown in the online Fig.~\ref{fig:mrlowmass}. It shows a single-peaked distribution centred around $q\sim4-5$, that is (given that we now examine systems with a primary mass $M_1=0.17$~M$_\odot$), $M_2=0.68-0.85$~M$_\odot$. In the figure, the corresponding mass-ratio distributions for the functional forms were computed with a single value of the primary mass. The outcome of the R-L method apparently agrees better with the Gaussian distribution, but given that we have now an even smaller sample, the data should not be overinterpreted
because the three distributions are again compatible within 2-$\sigma$. This peak may correspond to the similar peak seen in the mass-ratio distribution seen for the entire sample, while the fact that there is a possible second peak in Fig.~\ref{fig:massratio} at a lower mass ratio is most probably linked to the more massive primaries.

The very small excess of high mass ratios seen in Fig.~\ref{fig:massratio} and in the online Fig.~\ref{fig:mrlowmass} should not be given too much importance: firstly, it is a well-known effect of the R-L method to smooth the distribution, the effect becoming weaker depending on the number of iterations \citep{CB1994}, and secondly, their value is compatible with zero, given the size of the sample.

\section{Discussion and conclusions}

The results of \citet{ori} in the study of ELM WDs are very important,
therefore I have reanalysed the same sample of spectroscopic binaries they used with two different techniques. In the first one, I assumed functional forms and applied a K-S statistics. The obtained results were confirmed by a $\chi^2$ statistical test. The parameters for the Gaussian functional form are very similar to those found using a different method by \citet{ori}, but I  showed that a uniform distribution of the companion mass can provide as good a fit to the observed data and that the range of parameters allowed is rather large, making it hard to provide a definitive answer as to the real distribution. 
If the uniform distribution illustrated in Fig.~\ref{fig:distmass} is more representative of the companion mass distribution, then more double He WD binary systems may be expected, such as the eclipsing double white dwarf binary CSS 41177
 \citep{2014MNRAS.438.3399B}: the uniform distribution shown in this figure leads to a 24\% probability  (this is the fraction of systems that have a secondary mass lower than 0.5~M$_\odot$), compared to 16\% for the Gaussian fit shown in the same figure or as derived by \citet{ori}. If I take into account all the possible values at the 1-$\sigma$ level (i.e. those with $D^* < 0.1274$), I also derive a 26\% probability for both my Gaussian and uniform distributions.  
 
 In addition, I applied an inversion method to derive the mass-ratio distribution, without the need {\it \textup{to assume}} any functional form. The results are compatible with those derived from the functional form, although they seem to indicate some additional fine-grained structures.

An important question that \citet{ori} addressed is the possibility to have a neutron star as a companion to the ELM WD. \citet{ori} found that the probability of this is very low. This is also confirmed by my results. For example, the  uniform distribution of companion masses indicates that the range allowed for $D^*<0.0506$ ends at 1.4~M$_\odot$. If this is relaxed to $D^*<0.1274$, however,
the highest companion mass can be up to 1.7~M$_\odot$, but all in all, the probability to have one system with such a high mass ($> 1.44$~M$_\odot$) is very low: for the uniform distribution, the 1-$\sigma$ probability is zero, while for the Gaussian distribution it is 6.5\%.
 
\begin{acknowledgements}
I would like to thank the referee, J.J. Andrews, for a careful reading of the manuscript and for providing suggestions to improve the paper.
\end{acknowledgements}

\Online

\begin{figure}[htbp]
   \centering
   \includegraphics[width=9cm]{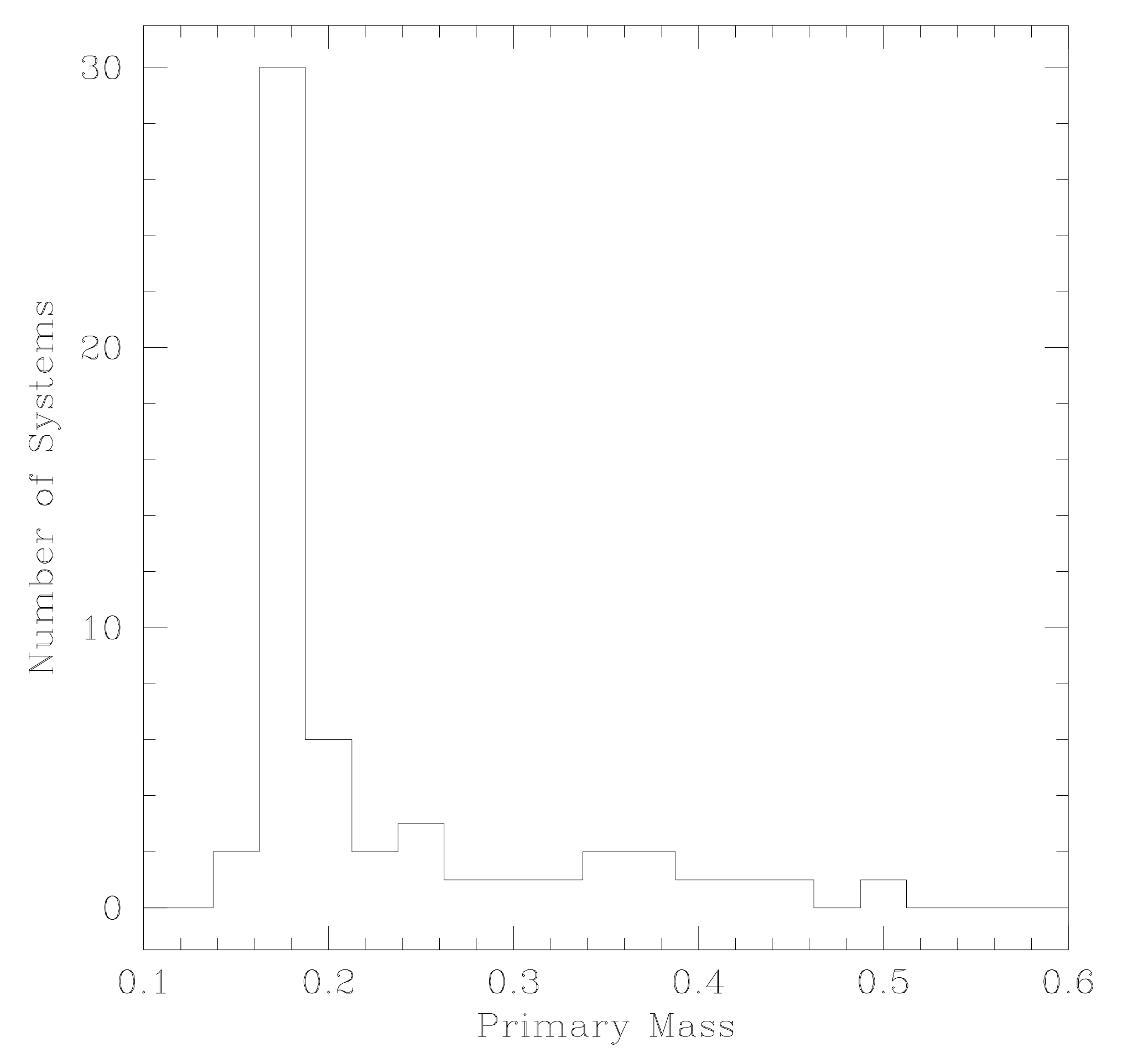} 
   \caption{Distribution of primary mass in the sample of \citet{Brown2013}. The sample contains many objects with primary mass equal to 0.17 M$_\odot$.}
   \label{fig:m1}
\end{figure}

\begin{figure}[htbp]
   \centering
   \includegraphics[width=9cm]{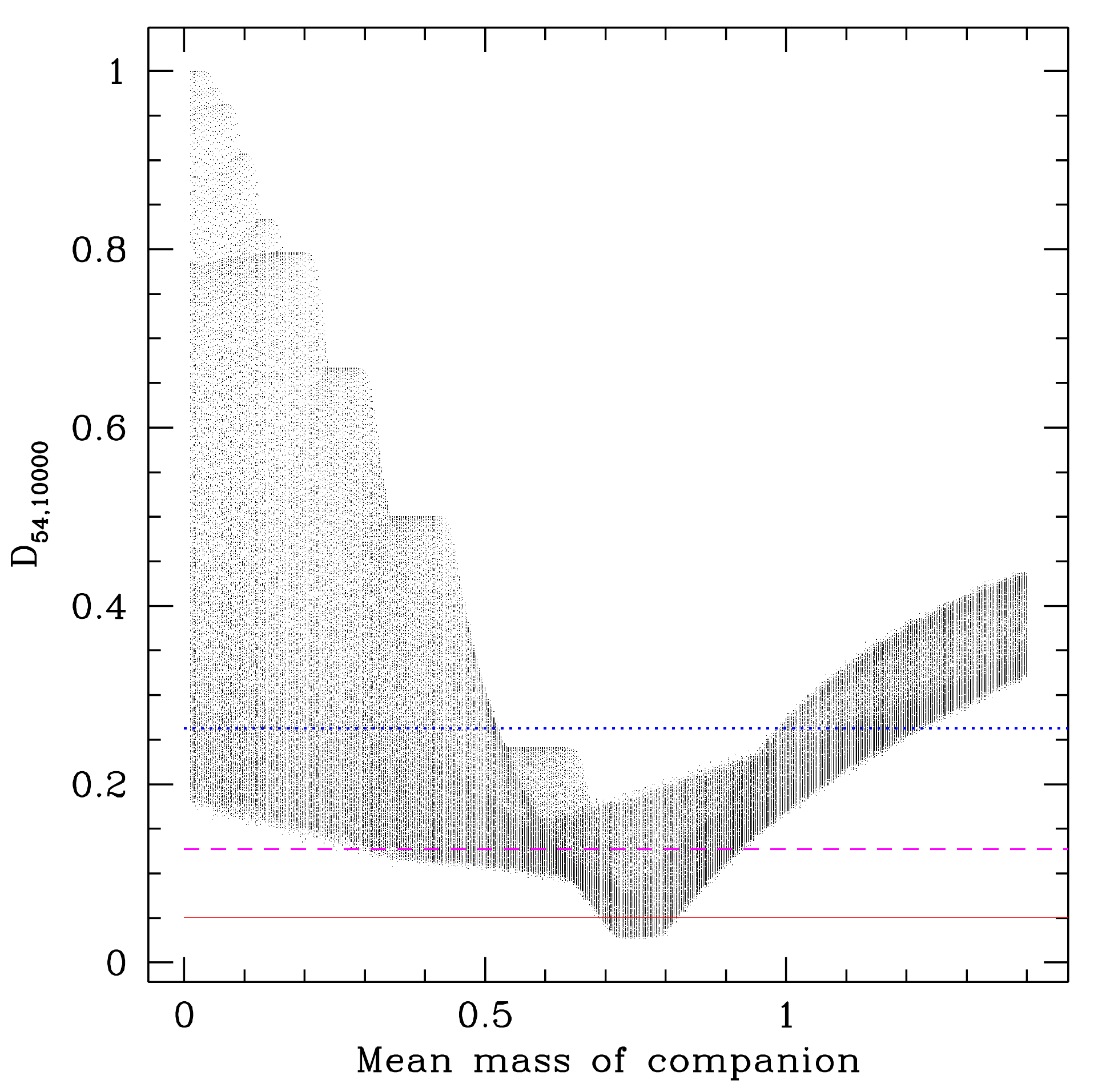} 
   \caption{Maximum deviation between the observed sample with $N=54$ and a synthetic sample containing $N^\prime=10,000$ objects, with the companion mass being distributed as a Gaussian with a mean mass as given by the abscissa. The blue dotted line is drawn at $D^*=0.2628$: all simulations with a higher value of the estimator  can be rejected at the 99.9\% level as being drawn from the same population as the observed distribution. The magenta dashed line is drawn at $D^*=0.1274$ and provide a 1-$\sigma$ estimate of the allowed parameter range, while the red solid line is drawn at $D^*=0.0506$.}
   \label{fig:gaussD}
\end{figure}

\begin{figure}[tbp]
   \centering
   \includegraphics[width=9cm]{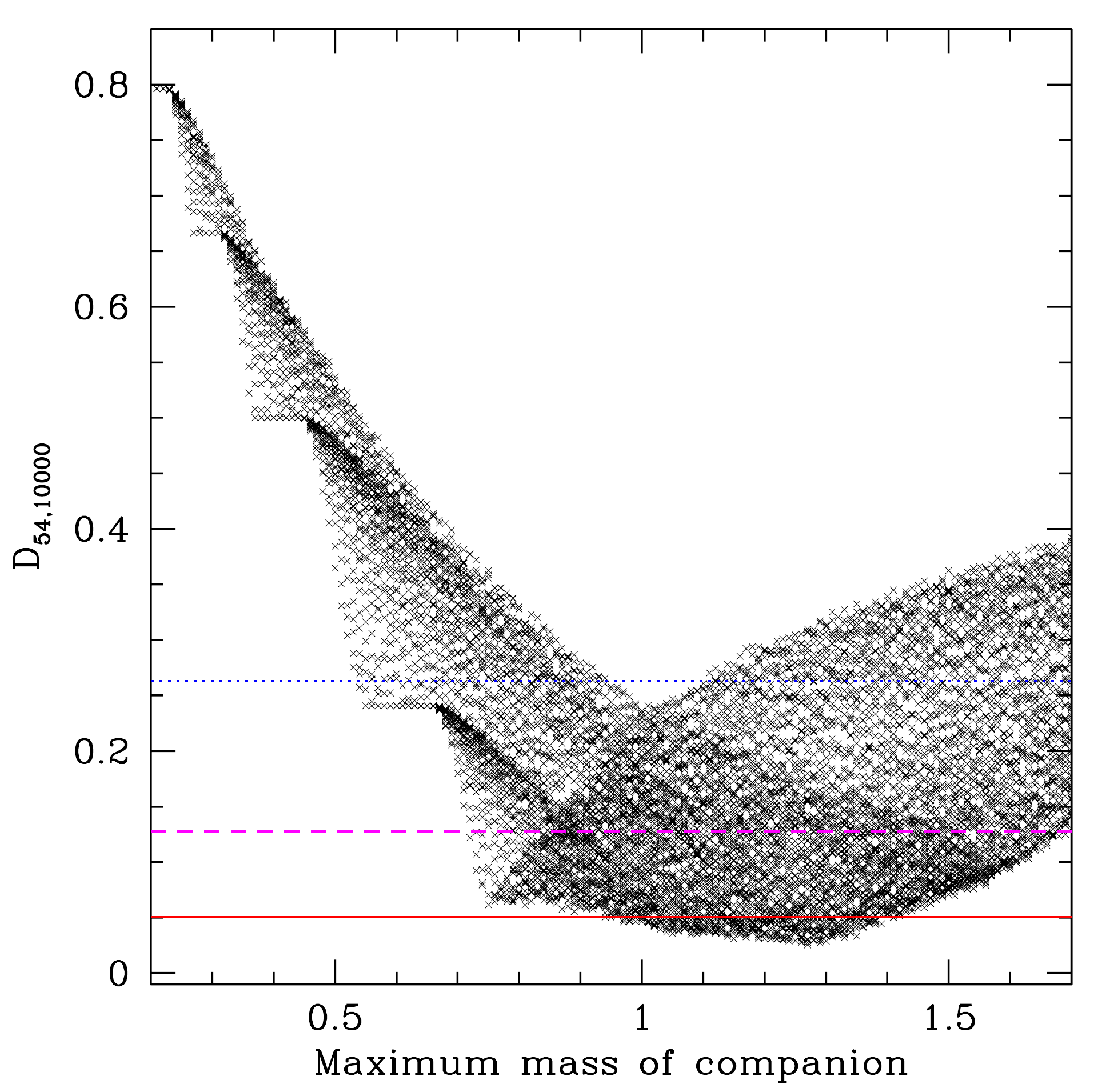} 
   \includegraphics[width=9cm]{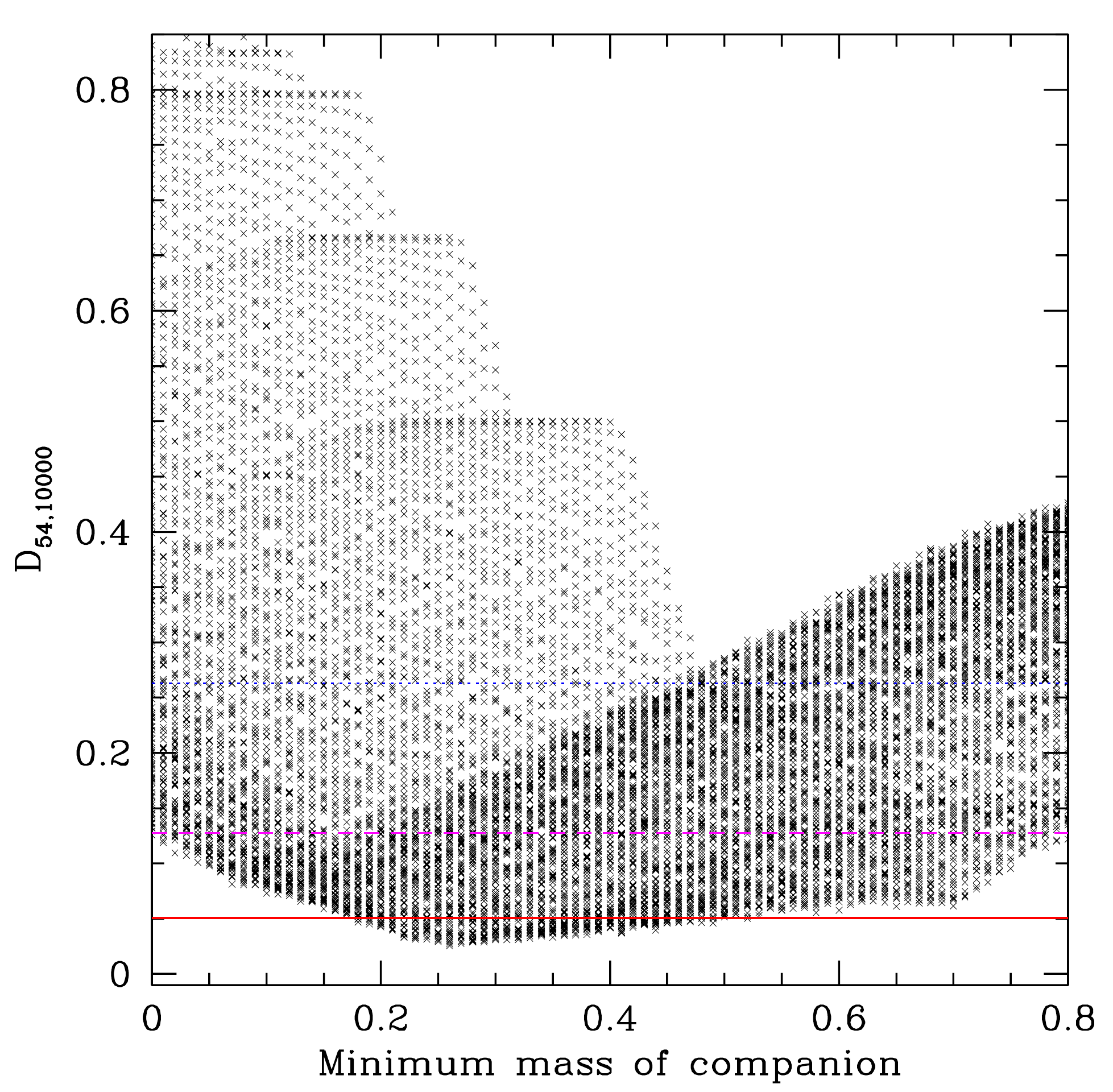}
   \caption{(top) Maximum deviation as a function of the highest mass of the companion if the latter is uniformly distributed between a lowest and highest mass. (bottom) The same, but as a function of the lowest mass. The lines are drawn at the same value of $D^*$ as in the online Fig.~\ref{fig:gaussD}.}
   \label{fig:uniD}
\end{figure}

\begin{figure}[htbp]
   \centering
   \includegraphics[width=9cm]{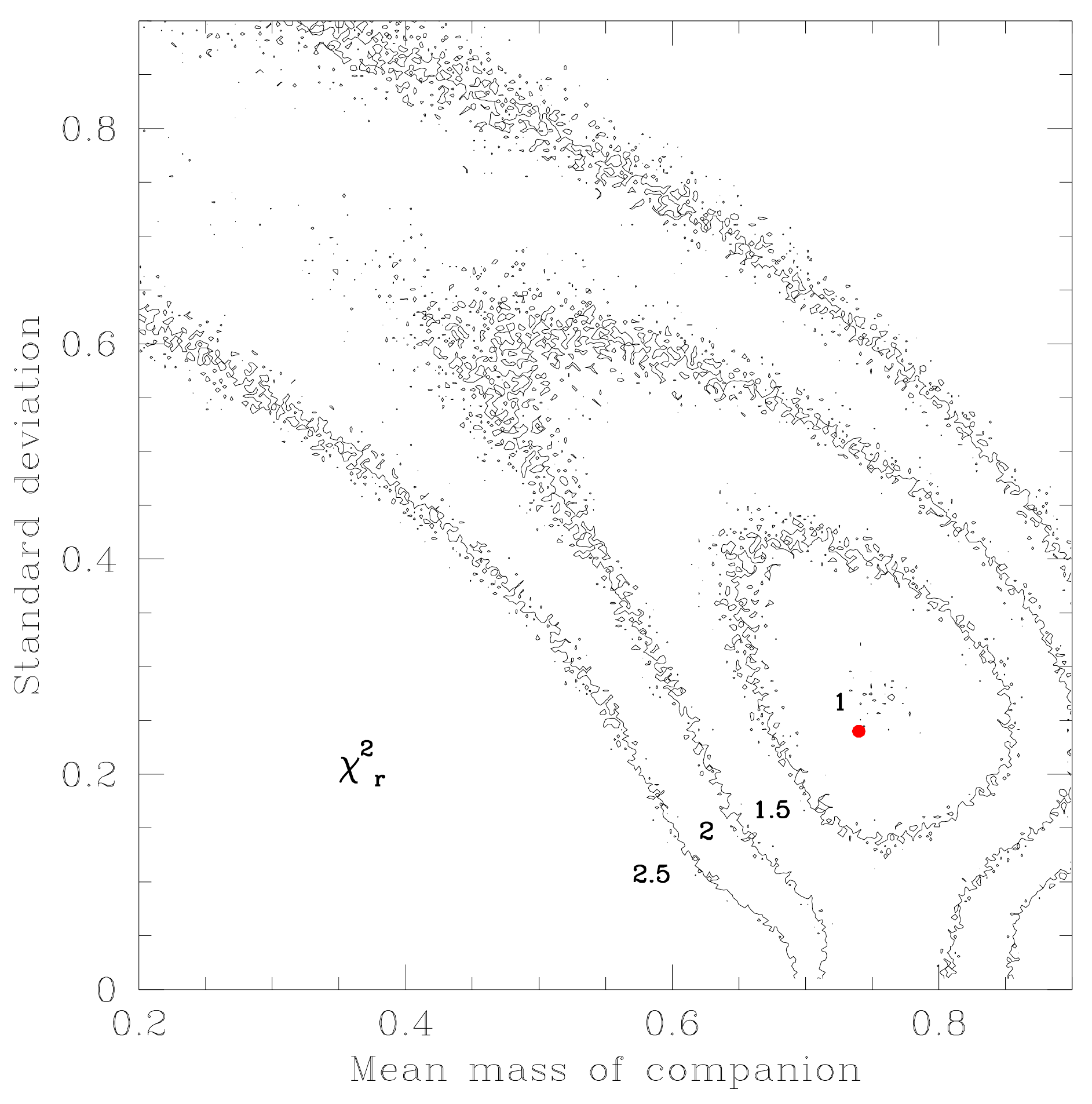} 
   \caption{Reduced $\chi^2$ map in the mean companion mass vs. standard deviation for an assumed Gaussian functional form of the companion mass distribution -- contours are shown at levels of 1, 1.5, 2, and 2.5. The red dot shows the position of the best fit of \citet{ori}.  The $\chi^2$ map confirms the result that was obtained with the K-S statistics.}
   \label{fig:x2map}
\end{figure}

\begin{figure}[tbp]
   \centering
   \includegraphics[width=9cm]{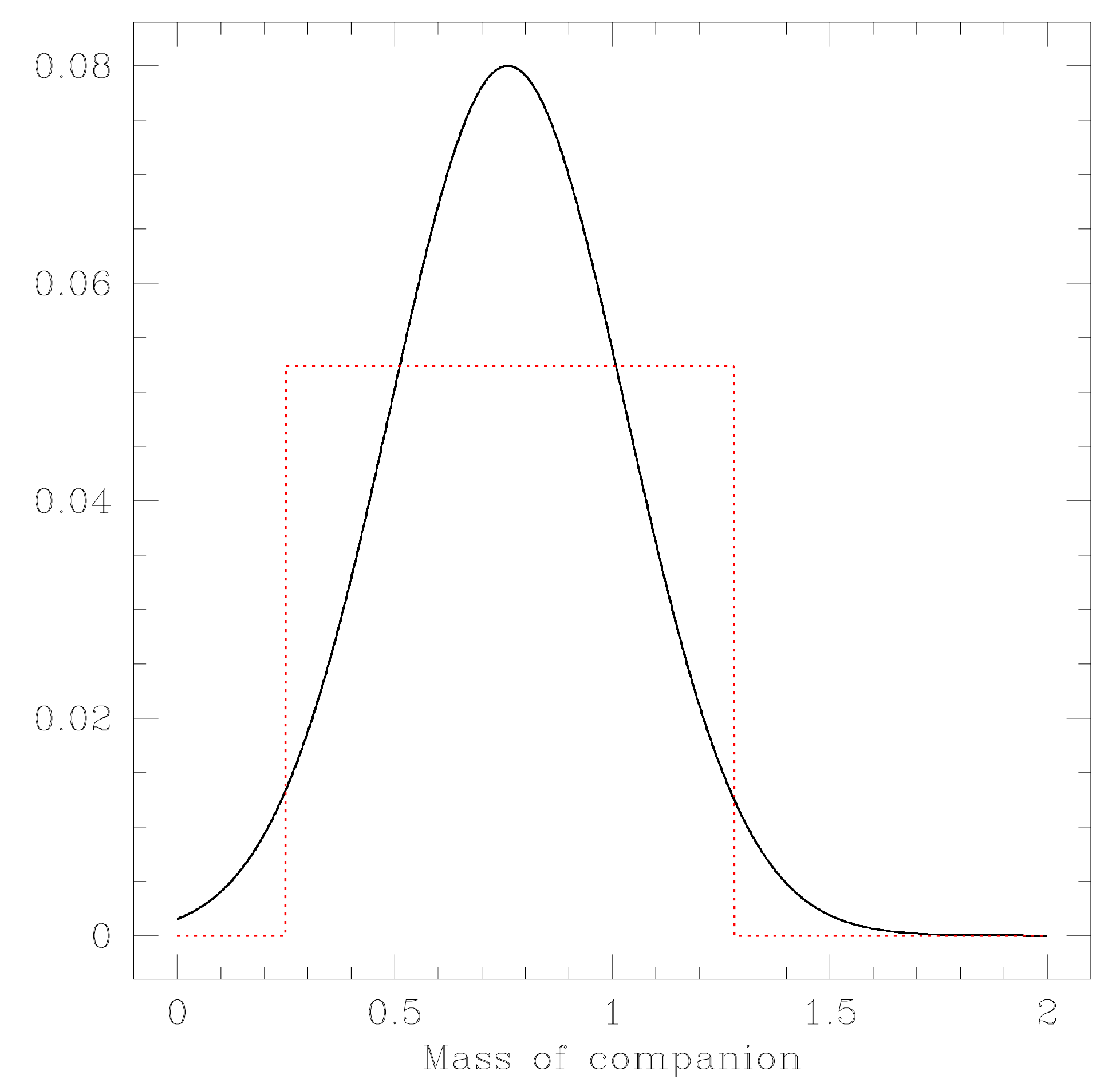} 
   \caption{Comparison between the two best functional forms that lead to the lowest values of $D^*$: a Gaussian distribution with $\mu = 0.76$~M$_\odot$ and $\sigma=0.27$~M$_\odot$ (solid black line) and a uniform distribution with $M_{2,\rm l}=0.25$~M$_\odot$ and $M_{2,\rm u}=1.28$~M$_\odot$ (dotted red line). }
   \label{fig:distmass}
\end{figure}

\begin{figure}[htbp]
   \centering
   \includegraphics[width=9cm]{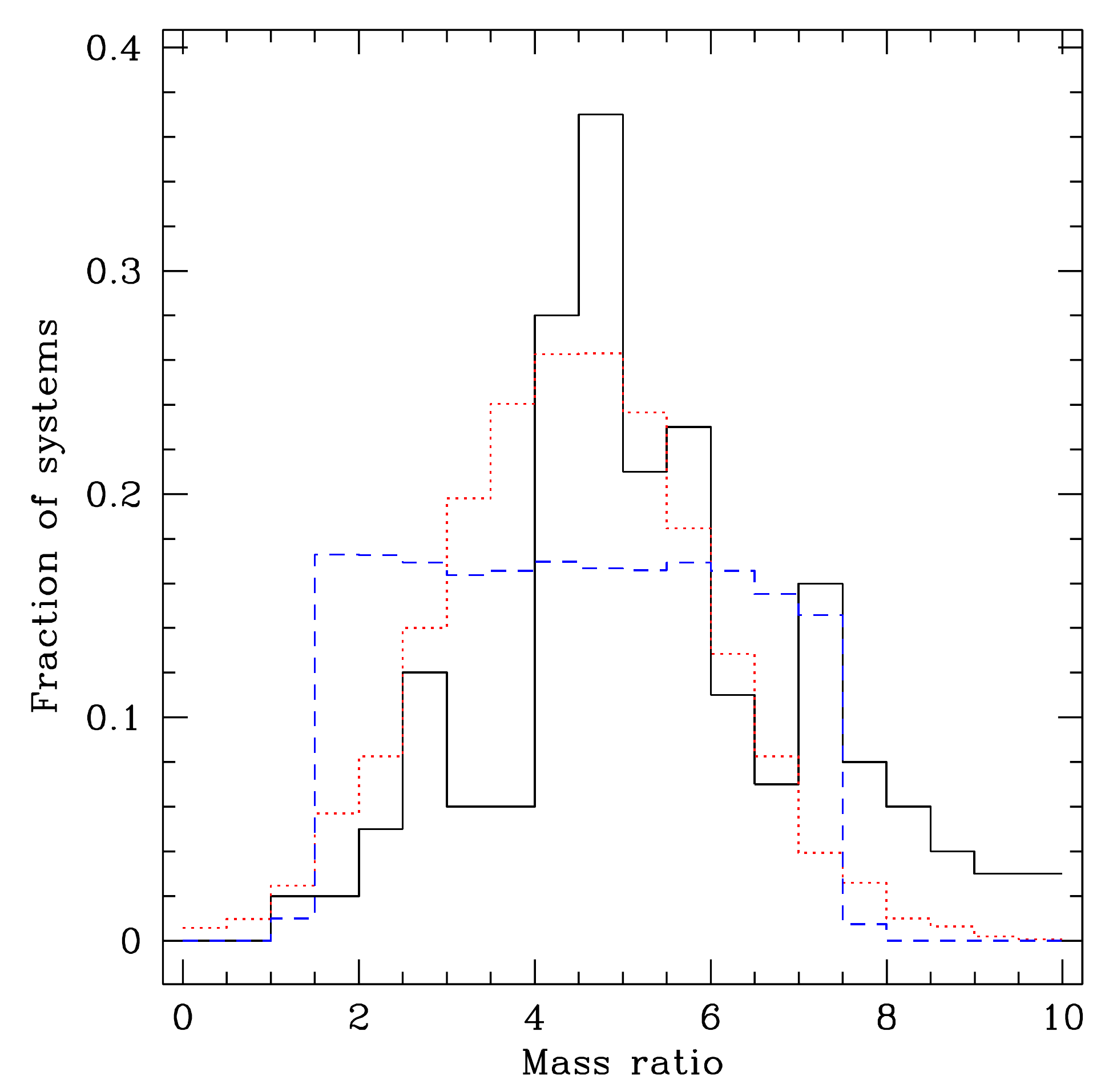} 
   \caption{Distribution of the mass ratios as determined by the Richardson-Lucy algorithm (solid black line) for all systems with a primary mass below 0.2~M$_\odot$, and as obtained for the best fits with a Gaussian  (red dotted line) and a uniform (blue dashed line) distribution of companions.}
   \label{fig:mrlowmass}
\end{figure}

\end{document}